# MITK-ModelFit: A generic open-source framework for model fits and their exploration in medical imaging – design, implementation and application on the example of DCE-MRI

Charlotte Debus[1-5,*,#], Ralf Floca[5,6,*,#], Michael Ingrisch[7], Ina Kompan[5,6], Klaus Maier-Hein[5,6,8], Amir Abdollahi[1-5], and Marco Nolden[6]

[1]*German Cancer Consortium (DKTK), Heidelberg, Germany*
[2]*Translational Radiation Oncology, German Cancer Research Center (DKFZ), Heidelberg, Germany*
[3]*Department of Radiation Oncology, Heidelberg Ion-Beam Therapy Center (HIT), Heidelberg University Hospital, Heidelberg, Germany*
[4]*National Center for Tumor Diseases (NCT), Heidelberg, Germany*
[5]*Heidelberg Institute of Radiation Oncology (HIRO), Germany*
[6]*Division of Medical Image Computing, German Cancer Research Center DKFZ, Germany*
[7]*Department of Radiology, University Hospital Munich, Ludwig-Maximilians-University Munich, Germany*
[8]*Section Pattern Recognition, Department of Radiation Oncology, Heidelberg University Hospital, Heidelberg, Germany*

# Correspondence:
Charlotte Debus, PhD
Department of Translational Radiation Oncology
Heidelberg Ion-Beam Therapy Center (HIT)
Im Neuenheimer Feld 450
69120 Heidelberg, Germany
Email: c.debus@dkfz-heidelberg.de
Phone: +49 6221 6538281

Ralf Floca, PhD
Division of Medical Image Computing
German Cancer Research Center (DKFZ)
Im Neuenheimer Feld 280
69120 Heidelberg, Germany
Email: r.floca@dkfz-heidelberg.de
Phone: + 49 6221 42 2560

---

[*] Shared first-authors

# Abstract

**Background:** Many medical imaging techniques utilize fitting approaches for quantitative parameter estimation and analysis. Common examples are pharmacokinetic modeling in dynamic contrast-enhanced (DCE) magnetic resonance imaging (MRI)/computed tomography (CT), apparent diffusion coefficient calculations and intravoxel incoherent motion modeling in diffusion-weighted MRI and Z-spectra analysis in chemical exchange saturation transfer MRI. Most available software tools are limited to a special purpose and do not allow for own developments and extensions. Furthermore, they are mostly designed as stand-alone solutions using external frameworks and thus cannot be easily incorporated natively in the analysis workflow.

**Results:** We present a framework for medical image fitting tasks that is included in the Medical Imaging Interaction Toolkit MITK, following a rigorous open-source, well-integrated and operating system independent policy. Software engineering-wise, the local models, the fitting infrastructure and the results representation are abstracted and thus can be easily adapted to any model fitting task on image data, independent of image modality or model. Several ready-to- use libraries for model fitting and use-cases, including fit evaluation and visualization, were implemented. Their embedding into MITK allows for easy data loading, pre- and post-processing and thus a natural inclusion of model fitting into an overarching workflow. As an example, we present a comprehensive set of plug-ins for the analysis of DCE MRI data, which we validated on existing and novel digital phantoms, yielding competitive deviations between fit and ground truth.

**Conclusions:** Providing a very flexible environment, our software mainly addresses developers of medical imaging software that includes model fitting algorithms and tools. Additionally, the framework is of high interest to users in the domain of perfusion MRI, as it offers feature-rich, freely available, validated tools to perform pharmacokinetic analysis on DCE MRI data, with both interactive and automated batch processing workflows.

*Keywords*: pharmacokinetic modeling, tracer-kinetics, dynamic PET, multi-purpose, software development, model fitting

# Background

Model fitting plays a vital role for many analysis approaches in medical imaging. In order to determine spatially resolved $T_1$ relaxation times in magnetic resonance imaging (MRI), multiple images with different $T_1$ weightings are acquired and the signal intensities are fitted with the relaxation equation [1]. Quantifying $T_1$ relaxation times can add additional morphological information for a variety of pathological conditions. In diffusion-weighted MRI (DWI), the apparent diffusion coefficient (ADC) is derived by acquiring images at increasing diffusion gradients (b-values) and fitting the signal loss with an exponential [2]. In more advanced signal theory, effects such as intravoxel incoherent motion (IVIM) are included, which also rely on signal fitting with a theoretical model. For chemical exchange saturation transfer (CEST) imaging, Z-spectra, acquired through sweeping radiofrequency saturation around the bulk water resonance, are analyzed using multi-pool Lorentzian fitting [3].

A paradigm for fitting of medical images is pharmacokinetic modeling, as applied in dynamic contrast-enhanced (DCE) MRI and computed tomography (CT), or in dynamic positron emission tomography (PET). In PET, pharmacokinetic analysis can be used to measure transport rates of certain pharmaceuticals or metabolic substances [4, 5]. Dynamic scans are acquired over the injection of a radioactive tracer, which accumulates in tissue according to the metabolic properties of its pharmacologic compound. Tissue-specific kinetic parameters are then extracted by fitting the measured time-activity curves with compartment models that describe tracer transport. The most commonly used models are the one tissue compartment model (1TCM) and two tissue compartment model (2TCM) [4].

In DCE MRI the aim is to derive parameters on tissue perfusion and capillary permeability from analysis of the time course of contrast agent (CA) concentration in tissue by acquisition of a time series of $T_1$ weighted MR images over CA administration [6, 7]. Tissue concentration-time curves are then analyzed through fitting with a pharmacokinetic (compartment) model [8, 9]. The most commonly used compartment models for gadolinium-based, extracellular contrast agents are the classical Tofts model, the extended Tofts model and the two compartment exchange model (2CXM) [6].

DCE MRI has become a popular method to assess tissue physiology in various diseases, including cancer, multiple sclerosis [10], rheumatic arthritis [11] and stroke [12]. For research purposes, authors usually write their own analysis code in general purpose frameworks like MATLAB [13], Python, R [14] or MPFit [15]. However, this approach comes with a number of disadvantages: in-house developed analysis software tools often lack standardization and broad validation, which can result in errors on the estimated parameter and make comparison of results from different centers rather difficult [16]. Also, code is often written without software design concepts and reusability in mind. Thus, novel applications or variations in the analysis workflow often have to be implemented from scratch. Furthermore, in-house developed tools usually lack the integration into medical image processing ecosystems, leading to excessive data conversion and transfer. This limits their application in clinical routine, as the fitting analysis cannot be performed directly together with other necessary data evaluation steps like segmentation and registration. Many times these in-house solutions are not graphical user interface (GUI)-based, and therefore require a basic knowledge of the respective programming language.

Due to these drawbacks, and especially with respect to standardization, clinically oriented studies tend to be carried out using standard scanner software tools included by the vendors (e.g [17, 18]) or stand-alone tools [19, 20]. Apart from their commercial nature, these tools constitute black-box systems that do not offer any flexibility in extension and configuration, which makes them less suitable for research purposes. Many of these tools offer only basic analysis steps and are installed on special workstations of scanner related computers. Hence data evaluation cannot be performed offline by any researcher. On top of this, studies have shown that results from different vendor's software yield differences in parameter estimates [21, 22, 23]. These aspects have given rise to the need of standardized, open access solutions.

**Challenges**

Ideally, such software tools would be included into larger medical image analysis platforms, enabling fitting analysis to be carried out side by side with other image processing steps without data conversion or import to other frameworks. In addition to that, linkage to a picture archiving and communication system (PACS) and support of DICOM data facilitates application of data evaluation in clinical settings. For research purposes, software should enable easy development and implementation of extensions to the tools in terms of models, fitting algorithms, etc. In order to be usable for both research and clinical evaluation purposes, the software needs to provide a user-friendly interface for analysis to be carried out, but yet allow for algorithm automatization in order to perform large scale data evaluations. Furthermore, direct means of fit visualization and exploration can improve quality of data evaluation and give room to model validation.

**State of the art – software**

Several open-source packages for analysis of DCE MRI data have been presented in the last years [24, 25, 26, 27, 28, 29, 30, 31]. They can be divided into two categories: stand-alone tools, designed explicitly and only for DCE MRI analysis, and plug-ins or packages, that provide the analysis functionality within a larger analysis framework. Stand-alone tools are designed to be ready-to-use applications, with ports for data transfer (data input and results export) and can be modified and extended on basis of the source code by the user. Well-known examples of stand-alone tools are ROCKETSHIP [24], DCE@urLAB [26] or DCEMRI.jl [27]. However, these tools are not linked to a common image processing platform, and thus require conversion and transfer of data. Hence, substantial effort is required to perform data analysis with these tools, making them feasible only for research purposes. Additionally, even though these tools are made publicly available as open-source code, many of them depend on some underlying closed source dependency, e.g. MATLAB including its toolboxes. Contrary to that, plug-ins or packages can be used within larger analysis frameworks. More general examples include published packages for R or Python, like DATforDCEMRI [25], dcemriS4 [32] or pydcemri [30].

More dedicated solutions were introduced to be used complementary to standard image processing software, like OsiriX [33], PMI [28] or 3DSlicer [34]. With regards to the aspect of clinical oriented analysis workflows, these plug-in solutions provide the advantage of incorporation of the DCE MRI analysis into general image pre- and post-processing. Thus, OsiriX plug-ins for DCE MRI analysis, e.g. UMMPerfusion [31] and the DCETool [29], are popular tools that can be used for an entire radiological workflow. However, OsiriX is only available on Mac OS, which presents another drawback. The 3DSlicer-based option "PKModeling" [35] on the other hand provides only basic features of pharmacokinetic analysis for DCE MRI data. Additionally, many of these plug-in solutions are designed for application with direct user interaction, thus not allowing for automated batch-processing analysis pipelines. Another aspect is that all the above named solutions are designed for a single application purpose, i.e. DCE MRI. In order to include other image processing tasks based on image fitting (especially on other fitting domains e.g frequency), would require entirely new implementations from scratch. However, general concepts of the underlying algorithm are not limited to DCE MRI. An ideal tool would offer means for fitting of medical image data with any model and on any domain (time, frequency, etc.)

To the best of our knowledge, there exists no solution, that can be considered truly free in terms of an open-source, operating system (OS)-independent software tool for fitting tasks on medical images, regardless of image modality, dimensionality and domain that does not depend in any way on external, commercial software frameworks. In this work, we present the framework ModelFit for the Medical Imaging Interaction Toolkit (MITK) [36], which is designed to perform any fitting task with a given model on multi-dimensional image data in such a free way. Several dedicated use-cases in form of MITK workbench applications were derived from this tool. Special attention was given to pharmacokinetic analysis in DCE MRI, for which several applications were implemented and validated.

# Implementation

We designed and implemented a framework within the Medical Imaging Interaction Toolkit (MITK) that enables fitting of medical imaging data with any given model. This framework was implemented with regards to both end-user applications as well as developer features. The following sub-sections present the design of the framework, explain how decoupling was achieved and which extension points are offered by the framework to tailor own setups and workflows.

## Definition of general terms and concepts:

Before we introduce the structure of the here presented framework, let us briefly review the conceptional workflow of data fitting. Data fitting is an optimization problem with the aim of approximating the measured data points by a theoretical mathematical model of the underlying (physical) processes.

The theoretical representation of the data is referred to as the model function $f_{\phi,\theta}(x)$ and maps from the signal grid domain $X$ to the signal codomain $Y$. $X$ and $Y$ are problem dependent. $X$ is e.g. often the time domain or frequency domain. $Y$ represents the intensities of the images (e.g. concentration of the CA). $X$ and $Y$ are subsets of $\mathbb{R}$.

The model function is parameterized by the parameter vectors $\phi$ and $\theta$. The parameter vector $\phi$ is the variable of the model fitting process and is named model parameter (MP) in the following. Parameter vector $\theta$ is not in the scope of the optimization and called static model parameter (SMP).

For the optimization the model $S'_{\vec{x},\theta}(\phi)$ in dependency of the signal grid $\vec{x}$ and the SMP $\theta$ is used. The values $S'$ are named signal (in analogy to the measured sample S). The input sample S is the vector of measured data points on $\vec{x}$. The optimization is performed by iteratively adjusting the set of MP in order to minimize a similarity measure between data and model, i.e. the deviation between sample and signal. This similarity measure is referred to as cost function $C(S, S'_{\vec{x},\theta}, \phi)$. $C$ may be a single scalar or vectorial. In many applications the sum of quadratic difference between the sample and theoretical signal, referred to as sum of squared residuals, is used as similarity measure.

## Decoupling strategies

An important design aspect for developers is the possibility to extend the framework in multiple ways and to reuse it for different fitting workflows and domains. Such a flexibility and reusability is achieved through the separation of concerns and decoupling (e.g. via abstraction). We regard these abstractions as equally important for a versatile fitting framework, though they are not sufficiently exploited in other publications (except for the model-view-controller pattern; see below).

1. **Abstraction of the model function**

Proper abstraction of the model function $f_{\phi,\theta}(x)$ seems trivial, but is nevertheless important for the versatility of the whole concept. The abstraction is done object-oriented via model classes that represent $S'_{\vec{x},\theta}(\phi)$, encapsulate the model function itself and generate signals for a defined signal grid upon request. Furthermore, a model class provides an abstract interface to interact with the encapsulated model function and to query its properties. The following properties are considered the most important regarding the fitting framework (e.g. for proper result serialization into DICOM and provenance tracking):

- Name/ID of the model
- Name and unit of the signal values
- Name and unit of the model parameters $\phi$ (MP) (i.e. parameters in the model function that are iteratively adjusted during fitting).
- Name and unit of the static model parameters $\theta$ (SMP) (i.e. parameters in the model function that are not affected by the fitting process)

## 2. Abstraction of the fitting process

The fitting process is abstracted into three components (see Fig. 1): model class (see above), cost function (e.g. sum of squared residuals; including the possibilities to define implicit regularization by boundary functions) and optimizer (e.g. LevenbergMarquart [37, 38] or LFGS-B [39, 40, 41]).

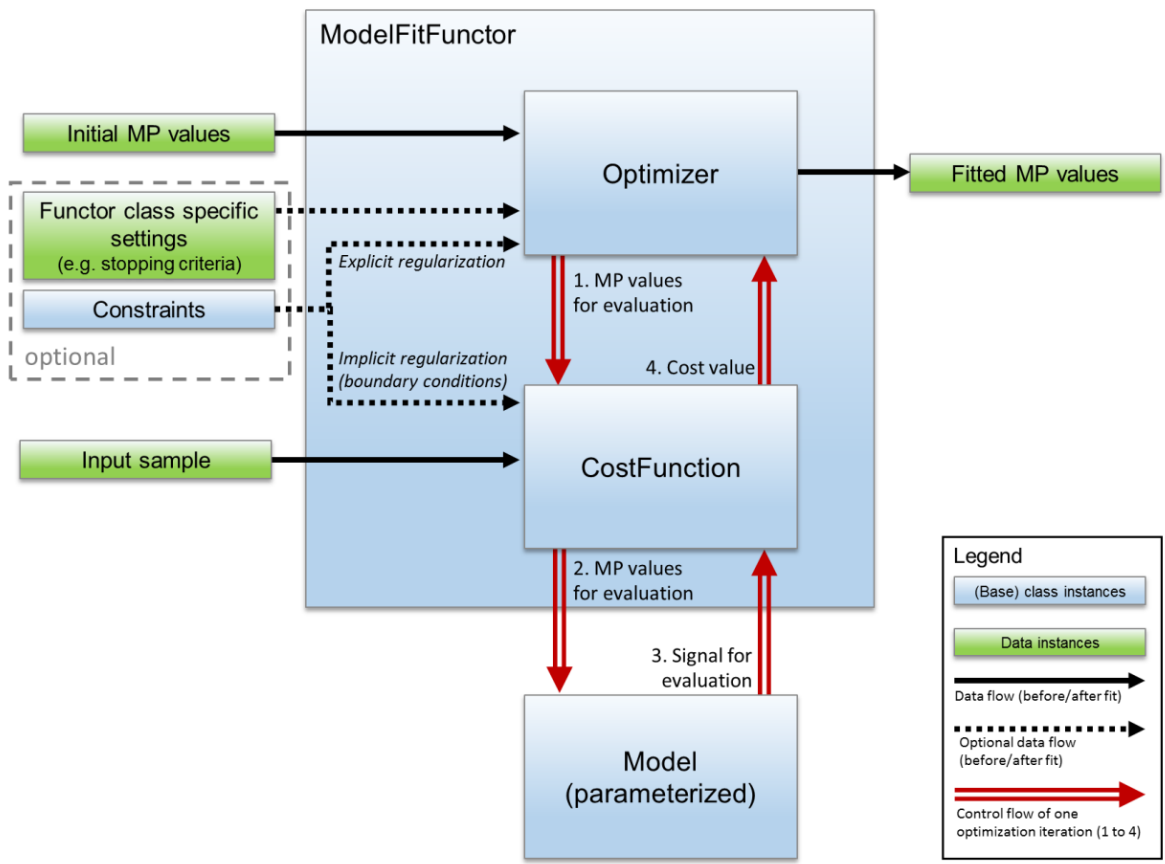

**Figure 1:** Abstraction of the fitting process. A ModelFitFunctor composes an optimizer and a suitable cost function. A ModelFitFunctor also depends on the input sample, initial MP values and a parameterized model instance. These are provided when calling the functor. Optionally a ModelFitFunctor class may specify additional settings (e.g. stopping criteria). Constraints may serve for explicit regularization (e.g. when using L-BFGS as optimizer) or for implicit regularization by boundary conditions that penalize the cost function. The control flow (red, double stroked arrows) of the optimization process loops through the steps 1 to 4 until a stopping criteria is met. Value class instances (green boxes) refer to input that is considered simple data. Base class instances (blue box) represent any derived class and are part of the abstraction

The optimizer drives and terminates the iterative fitting process based on the cost function and the optimizer's stopping criteria. The combination of optimizer, cost function and model can be arbitrary chosen by the developer for the desired fitting workflow. Amongst others aspects, this allows for experimental settings e.g. benchmarking the performance of different model implementations using the same optimizer and cost function [42]. To facilitate reuse, meaningful combinations of optimizers and cost functions are composed as ModelFitFunctor classes. ModelFitFunctors depend only on MPs, the parameterized model itself and the sample. This design allows a great versatility and reusability in different workflows; e.g. the ModelFitFunctor itself does not change, if either a region of interest (ROI)-based (averaged curve over all voxels in a region of interest) or voxel-based fit (each voxel individually) is performed. To achieve this versatility, the fitting process (Fig. 1) is completely abstracted from type or purpose of input and output data. The concrete realization and further benefits are explained in more details in the next section.

## 3. Abstraction of data

To conduct fitting, several types of information are needed:

- Sample signal
- SMPs
- Initial MP values
- Constraints for the fitting

With regard towards fitting workflows and the above mentioned information types the following design consideration was made: Fitting is always done for an indexed discrete element (e.g. an image voxel). Therefore any data can be defined on a global scope (e.g. the sample signal in a ROI-based fitting) or a local scope (e.g. the sample signal in a voxel-based fitting; initial MP values of a model). The type of scope is not fixed. It might change depending on the chosen model, the workflow and the experimental setting. Furthermore the source of data and its representation (values stored in an image object, a value vector, etc.) might differ, depending on the workflow and state of the application.

In the here presented framework, this consideration is dealt with by introducing two groups of classes: ModelParameterizer classes and ModelFactory classes. The interplay of these classes with the fitting process is depicted in Fig. 2. The ModelParameterizer abstracts the way how default constraints, initial MP values and SMPs are accessed and therefore, the handling of different data representations and scopes. The ModelFitFunctor uses the ModelParameterizer for any index to request a parameterized and ready-to-use model instance with initial MP values for fitting.

ModelFactory classes are used by the application to get a valid ModelParameterizer based on the application state and available data types. Hence a ModelFactory encapsulates the decision, which ModelParameterizer should be used and how it should be initialized. A ModelFactory always represent a model class in the context of a certain problem statement. Thus, one model class might be managed by several ModelFactories, but with different ModelParameterizers and constraints regarding the specific problem statement, for which the factory was implemented.

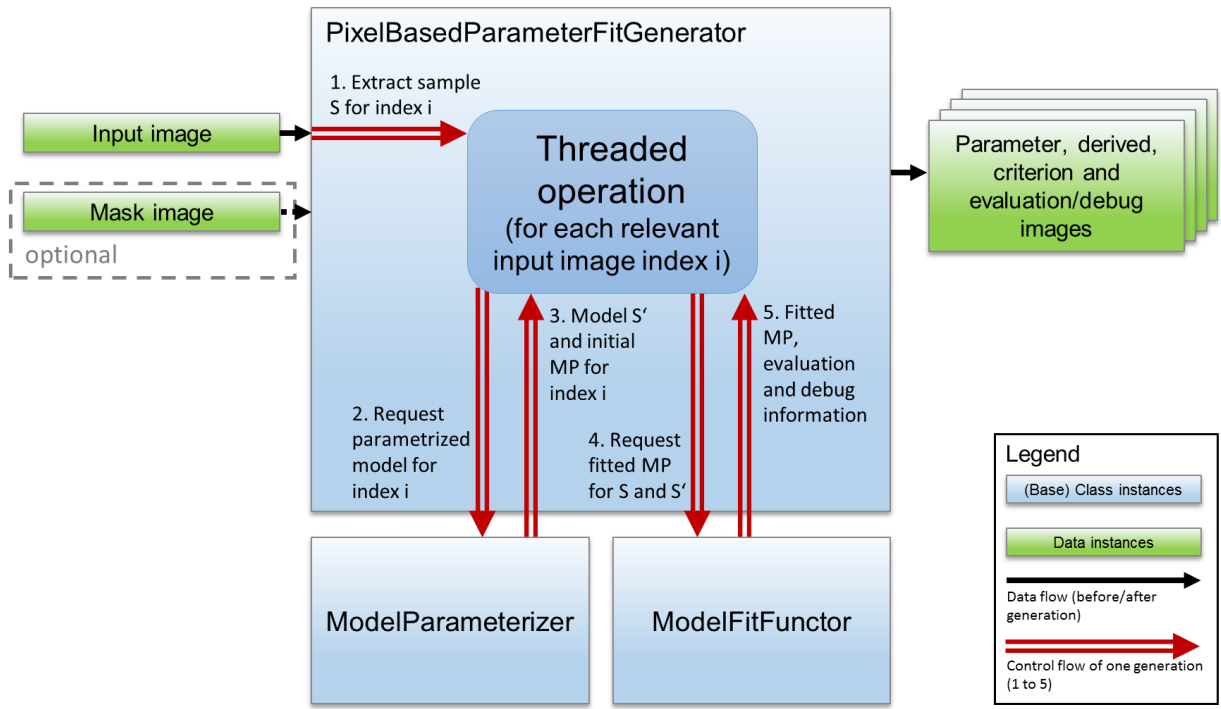

**Figure 2:** Illustration of the fitting process using the example of voxel-wise fitting. The PixelBasedParameterFitGenerator computes the fits concurrently for all relevant voxels (identified by the optional mask). The generator interacts with a ModelParameterizer and a ModelFitFunctor instance that should be used for the generation. The control flow (red, double stroked arrows) of the generation process loops through the step 1 to 5 for each voxel index. Output is, besides the parameter images, a criterion image (representing the final cost function value of the fit), evaluation maps (representing additional user defined measures for fit quality) and optional debug images (containing ModelFitFuctor specific information like number of iteration or met stop criterion of a fit). Value class instances (green boxes) refer to input/output that is considered simple data. Base class instances (blue box) represent any derived class and are part of the abstraction

### 4. Model-View-Controller pattern

The model-view-controller (MVC) pattern [43] and its variations are well-known strategies to decouple parts of an application and to allow thorough separation of concerns. It has been applied in other solutions [44]. In our implementations based on MITK, a MVC pattern with multiple views and controllers was applied. To avoid the ambiguity of the term "model" in the context of this paper, the term "application model" will be used for the model of the MVC pattern. In all other cases "model" refers to model classes that represent $S'_{\vec{x},\theta}(\phi)$ (see above).

In the herein presented framework, the application model not only consists of the data (e.g. input images, ROIs, resulting parameter images) but also of the fitting business logic. The fitting business logic encompasses all classes and structures introduced in the above abstraction levels (e.g. model class, ModelFitFunctor classes, etc.). The decision to make the fitting business logic part of the application model instead of the controller allows its decoupling from controllers. This decoupling enables the reuse of fitting business logic components in multiple controllers and facilitates the necessary unit testing. Within the MITK workbench implementation, a view consists of multiple graphical user elements (widgets) that display the images, model functions, model constraints etc. The controllers are provided by MITK workbench plug-ins (MFI, generator plug-ins for DCE MRI, etc.).

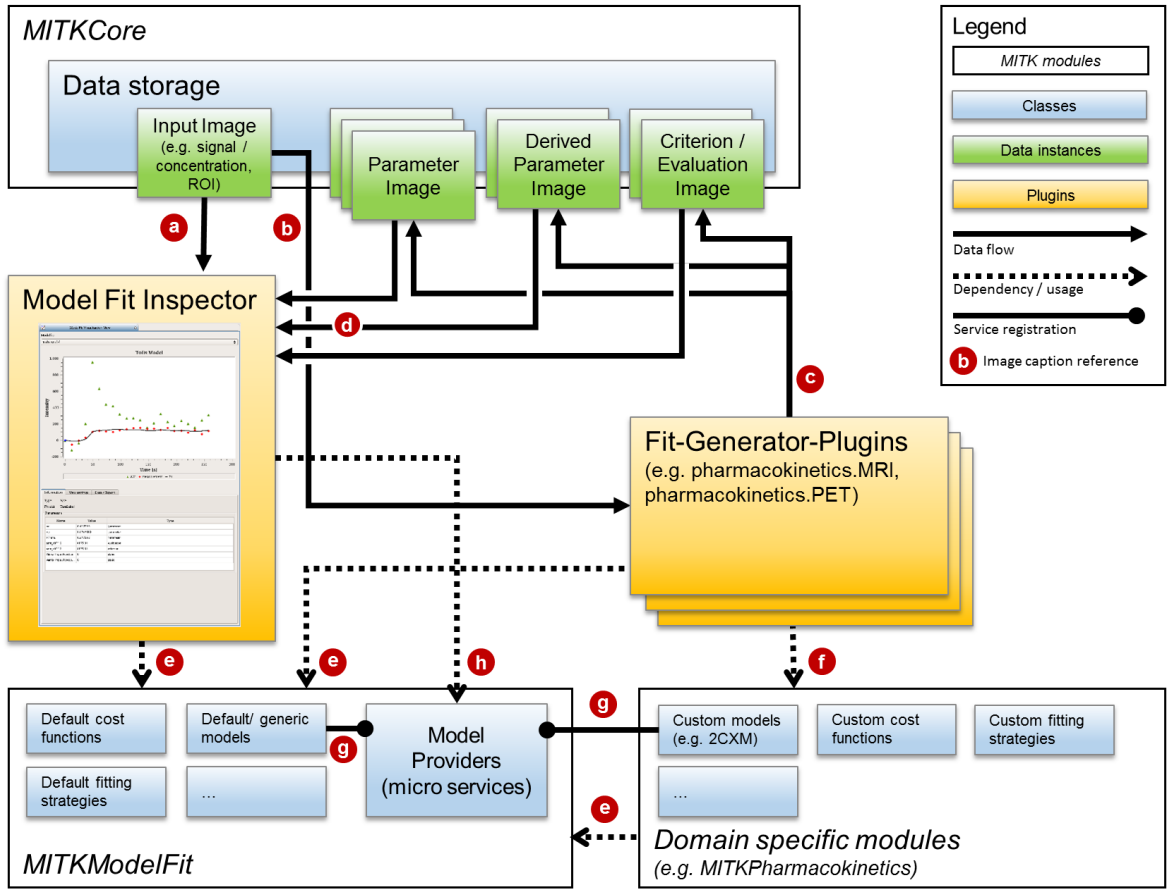

**Figure 3:** Simplified illustration of the interplay between components for model fitting. The plug-ins (yellow boxes) represent the MVC controllers. Data (green boxes) are part of the MVC application model together with the modeling relevant classes (blue boxes; bottom part). The Model Fit Inspector visualizes raw 3D+t input data (**a**) and, if present in the data storage, uses the result of fits (**d**) to visualize the fits. Fits are generated by domain specific generator plug-ins that use the inputs (**b**) and store the results (**c**) in the data storage. The whole fit information is encoded in the result images and their meta information. All fitting plug-ins and domain specific modules (e.g. pharmacokinetics) depend on parts of the ModelFit module (**e**). In addition, domain-specific plug-ins also depend on specific modules (**f**) that provide the model, cost function or ModelFitFunctor classes of the domain. To allow every part of the application to use a specific model class, they are registered (**g**) by their modules via micro services (model provider). The model providers are e.g. used (**h**) by the Model Fit Inspector to plot the respective model signals without establishing code dependence on any generator plug-in or domain module

The application model is decoupled from the views and the controllers in two ways: data is decoupled via the MITK data storage and the MITK data / properties classes, which grant access to data and its meta data. Hence controllers and views do not interact directly, but via the information in the MITK data storage and application events. To decouple the model business logic from the controllers micro services are used to inject ModelFactory classes into the application and allow arbitrary controllers to access them. The MVC pattern of our application and its interplay is depicted in Fig. 3.

## Extension points of the framework

As an open-source project there is a vast variety of options to extend and customize the described framework. Five extension points are regarded as most important and will therefore be explained briefly.

- **New models:** The introduction of new models is the most obvious one. A completely new model function requires also the implementation of a respective ModelParameterizer and ModelFactory class. For both types, template base classes are provided to facilitate implementation. If a developer wants to add support of different data representations for an existing model class, only a suitable ModelParameterizer and ModelFactory must be implemented. For the registration of a factory as a micro service in order to make it available in the application, a helper class is provided.
- **New cost functions and fitting strategies:** Custom cost functions can be integrated based on two base classes: SVModelFitCostFunction is used for single-value cost functions (e.g. sum of squared residuals) and MVModelfitCostFunction for multi-value cost functions (e.g. array of squared residuals). Both classes are based on cost function classes of the Insight Toolkit (ITK)[45], itk::SingleValuedCostFunction and itk::MultipleValuedCostFunction respectively. Therefore, every thread-safe optimizer offered by ITK can be used to drive the fitting process. In addition, own implementations or wrapping of existent optimizer implementations are possible. In order to regard different types of fitting constraints and boundary conditions, an abstract interface is provided along with a ready-to-use implementation of simple boundary conditions. The interface can either be used to inject constraints implicitly or explicitly into the fitting process. The latter option must be supported by the optimizer itself (e.g. LBFGS-B). The implicit injection is realized by a penalty term added to the cost function. This is easily done by using SVConstrainedCostFunctionDecorator or MVConstrainedCostFunctionDecorator. Both take the constraints and the original cost function and can then be used as cost functions with penalty term.
- **New domains:** The main area of usage is currently time-resolved data, e.g. in pharmacokinetic analysis or for simple trend fitting (e.g. linear or exponential fits). Nevertheless, the framework itself is not limited to a special domain, neither data specific (e.g. time domain or frequency domain) nor regarding the use-case (e.g. image modalities, types of models). This covers e.g. the fitting of diffusion data over different b-values for ADC extraction, the determination of $T_1$ over different inversion times and flip angles or $T_2$* fitting for variant echo times. Due to the above introduced abstraction levels the only restriction imposed by the framework is the possibility to implement the model function. Everything else is covered by ready-to-use classes or can be extended.
- **New controller/generation plug-ins:** At the latest when adding a new domain to MITK GUI applications, one has to add a new generation plug-in to serve as a controller. To ease the implementation of custom generation plug-ins, many typical generation aspects are encapsulated into ready-to-use widgets (e.g. definition of initial values or definition of constraints). Due to the above introduced abstractions those widgets can work with any model. Therefore developers can concentrate on implementing the logic that initializes the needed ModelFitFunctor.

# Results

Due to the presented used decoupling strategies, the fitting framework is very flexible in terms of which use-case can be addressed and how the use-case is implemented. The first aspect (what/which) is possible due to abstraction between model, data and fit representation (see abstraction strategy 1 and 3). Thus, the framework can be applied to any kind of fitting task regardless of the image modality (CT, MRI, PET, etc.), fitting dimension (time, frequency, etc.) and applied model (linear, pharmacokinetic, etc.).

The second aspect (how) is achieved by separation and standardization of the fitting routine components in terms of model, cost function (respective fitting criteria) and fitting engine (optimizer) as well as by the used MVC pattern (see abstraction strategy 1, 2 and 4). This leads to modularity and high flexibility for implementations of concrete fitting workflows and applications. The versatility is demonstrated by the implementation of several ready-to-use tools, which will be shortly presented in the following

There are several different solutions, including our own work, available for the fitting of medical data (especially for DCE MRI). To ease the comparison and assessment for developers and users, Table 1 compiles software characteristics for a selection of solutions. The selection of solutions represents well-known or relatively similar solutions compared to our work, in order to show differences between potential alternatives. The selection is not exhaustive.

For exploration of dynamic data and respective fits, the Model Fit Inspector (MFI) allows voxel-wise display of multi-dimensional data and associated fits. If no model is fitted to the data, it displays the raw image intensity values over time in the selected voxel. The plug-in can be used to scout the data (see Fig. 4), visually assessing data quality (temporal sampling, noise, etc.) and qualitatively evaluate the course of signal-time curves. After fitting, the resulting fit curve can be displayed together with the measured intensity time curve it was fitted to. Data properties like noise or different curve shapes can be assessed visually by navigating through the image. For ROI-based fits, both averaged curve and curve at the specific image position are shown. If an additional curve was defined (e.g. as input for the model, such as an arterial input function (AIF)), it is displayed as well in a different color. Display settings (axis range, curve display color) can be adjusted manually. An info box shows all resulting parameter estimates, evaluation and derived parameters for that specific fit. Data curves can be exported as [Time, Signal] arrays for external analysis.

For fitting tasks outside of pharmacokinetic analysis we provide a simple tool. Currently this multi-purpose fitting tool offers conduction of simple fits e.g. with linear or exponential models as well as a generic model. The generic model uses a formula parser to fit any explicit analytical model formula to data. The user needs to specify the functional representation of the model and the number of model parameters that are adjusted during fitting.

When data quality is not sufficient to enable proper fitting analysis or no suitable model is known, simple, semi-quantitative parameters describing the curve shape can be calculated to evaluate the data [46, 47, 48].

|  | MITK- ModelFit | UMM Perfusion | Rocketship | DCEMRI.jl | PMI | DATforDCEMRI | 3DSlicer PkModeling |
|---|---|---|---|---|---|---|---|
| Operating system | Linux, Mac OS, Windows | Mac OS | Linux, Mac OS, Windows | Linux, Mac OS, Windows | Windows | Linux, Mac OS, Windows | Linux, Mac OS, Windows |
| Language | C++ | C | Matlab | Julia | IDL | R | C++ |
| License | BSD | BSD | GNUGPL | MIT | GNU GPL | Creative Commons | Slicer (BSD like) |
| Advanced extensibility | Yes | Yes | No | No | No | No | Yes |
| Fitting domain | Time, Frequency, any* | Time | Time | Time | Time | Time | Time |
| Eco-system | Yes (MITK) | Yes (OsiriX) | No | No | Yes (PMI) | No | Yes (3DSlicer) |
| Image modalities | DCE-MRI, DCE-CT, PET, dynamic MRI, dynamic CT, CESL/CEST, * | DCE-MRI | DCE-MRI | DCE-MRI | DCE-MRI, DSC-MRI, DCE-CT | DCE-MRI | DCE-MRI |
| Models | Tofts, Extended Tofts, 2CXM, 1TCM, 2TCM, Brix, Three-step linear (3SL), Semi-quantitative metrics (BAT, TTP, AUC, $C_{max}$, Wash-in/Wash-out Slope, final uptake, mean residence time) | Extended Tofts, 1CP, 2CXM, 2C uptake model, two compartment filtration model (2FM) | Tofts, Extended Tofts, Fast Exchange Regime, 2CXM, Tissue uptake, Nested-model selection, Patlak, Semi-quantitative metrics (AUC) | Tofts, Extended Tofts, Plasma Only | Uptake models, Steady-state, Patlak, Model-free deconvolution, Tofts, Extended Tofts, 2CXM, 2C filtration model for kidney, Dual-inlet models for Liver, Semi-quantitative metrics (Slope/Signal enhancement) | Tofts, Semi-quantitative metrics (AUC, MRT - mean residence time) | Tofts, Semi-quantitative metrics (AUC, slope) |
| Input / Output | DICOM, Analyze, NIFTI, NRRD, VTK, Raw data | DICOM | DICOM, Analyze, NIFTI, Raw data, Matlab data | Matlab data | DICOM, Raw data | R readable data formats | DICOM, Analyze, NIFTI, NRRD, VTK, Raw data |
| GUI | Yes | Yes | Yes | No | Yes | No | Yes |
| Fit exploration | Yes | Yes | Yes | No | Yes | No | Yes** |
| PACS Support | Yes | Yes | No | No | No | No | Yes |
| Automatization | Yes | Partially*** | Yes | Yes | Yes | Yes | Yes |
| Source | https://www.mitk.org | http://ikrsrv1.medma.uni-heidelberg.de/redmine/projects/ummperfusion | https://github.com/petmri/ROCKETSHIP | https://github.com/davidssmith/DCEMRI.jl | https://sites.google.com/site/plaresmedima/ | https://github.com/cran/DATforDCEMRI | https://www.slicer.org/wiki/Documentation/4.8/Modules/PkModeling |

\*: Possibility to extend framework to support other fitting domains.

\*\*: Possibility to generate a 3D+t image that encode the voxel-wise model signal and to explore the image with the MultiVolumeExplorer.

\*\*\*: Possibility to loop over all models and selected tissue ROIs for the loaded Data in the UMMPerfusion user interface.

**Table 1**: Software characteristics. The selection of solutions represents well-known or relative similar solutions compared to our work in order to clarify the differences. The selection does not claim to be exhaustive. Commercial solutions are not included. Further R or Matlab are only included in context of concrete tools (DATforDCEMRI and Rocketship) and not as generic fitting environments on their own. The later would be a categorical error. R as well as Matlab can handle generic fitting problems or allow GUIs but by implementing an application from scratch and not by just using it of the shelf or extending an existing one.

The following characteristics are assessed in the table: Operating system; Language (Programming language of the software); License (needed to regard if software is used/extended); Advanced extensibility (Indicates if software was designed to easily be extended with new models without the need to change the basis application or its programming logic; implies a advanced level of abstraction and decoupling); Fitting domain (Indicates which domains are supported for the fitting); Eco-system (indicates if software is embedded into image processing eco-system); Image modalities (medical image modalities that are supported be model and fitting techniques); Models (included pharmacokinetic models); Input / Output (most relevant data formats supported by the software); GUI (indicates if software offers a graphical user interface); Fit exploration (indicates if the software allows to interactively investigate the fit and signal curve per voxel/ROI); PACS Support (indicates if the software allows to use DICOM Q/R or receive data via DICOM Send); Automatization (indicates if the software can be used to automatize the analysis with no user interaction); Source (Link to the source codes or developer's site).

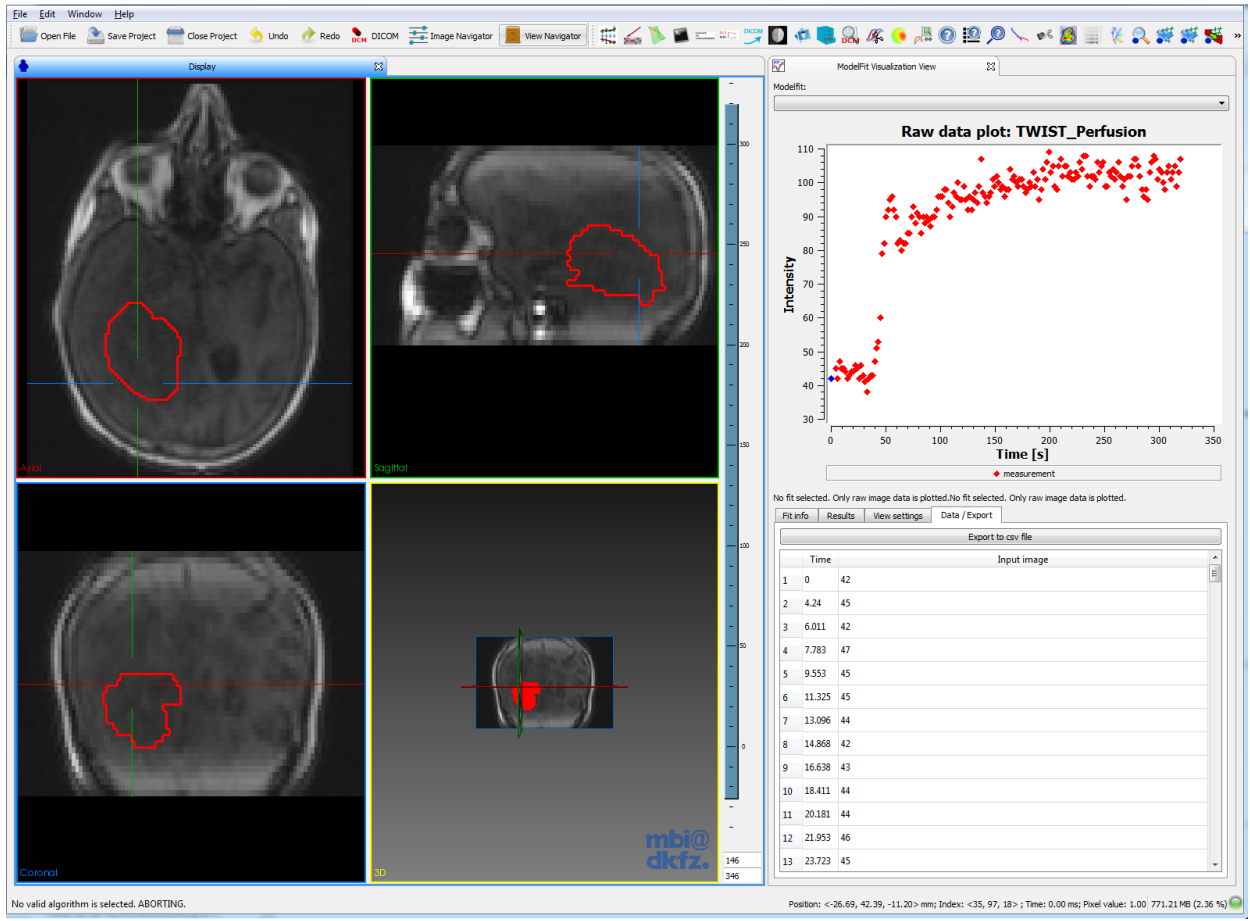

**Figure 4:** Screenshot of the MITK workbench and the MFI plug-in (right), with exemplary DCE MRI data from a glioblastoma patient. In the 4-window view, the acquired 3D images can be viewed at each time point. The MFI plug-in enables display of the signal-time curves in each image voxel (crosshair). The respective signal-time curve can be exported as 2-column .csv file

For these cases a plug-in for non-compartmental analysis of signal-time curves using semi-quantitative parameters (depicted in Fig. 5) is provided. Common examples are the integral area-under-the-curve (AUC), the maximum signal intensity or time-to-peak (TTP). In pharmacokinetic theory, this approach is often referred to as non-compartmental or descriptive analysis. The set of parameters is extendable and currently includes AUC, maximum intensity, TTP, area-under-the-first-moment curve and mean residence time [49]. The resulting parameter images can be further analyzed or used to identify regions of interest for detailed pharmacokinetic analysis.

DCE MRI data can be quantitatively analyzed with pharmacokinetic models using the DCE MRI fitting plug-in. It includes a descriptive model [50], the classical Tofts model, the extended Tofts model and the two compartment exchange model (2CXM). The 2CXM is provided in both the convolution and differential equation form [42]. Figure 6 shows an example for pharmacokinetic analysis, in this case DCE MRI data from a glioblastoma patient that was analyzed using the 2CXM. Furthermore, a simple three-step linear model was implemented, that assumes linear functions for each three segments of the curve in order to derive semi-quantitative measures, like the wash-in or wash-out slope.

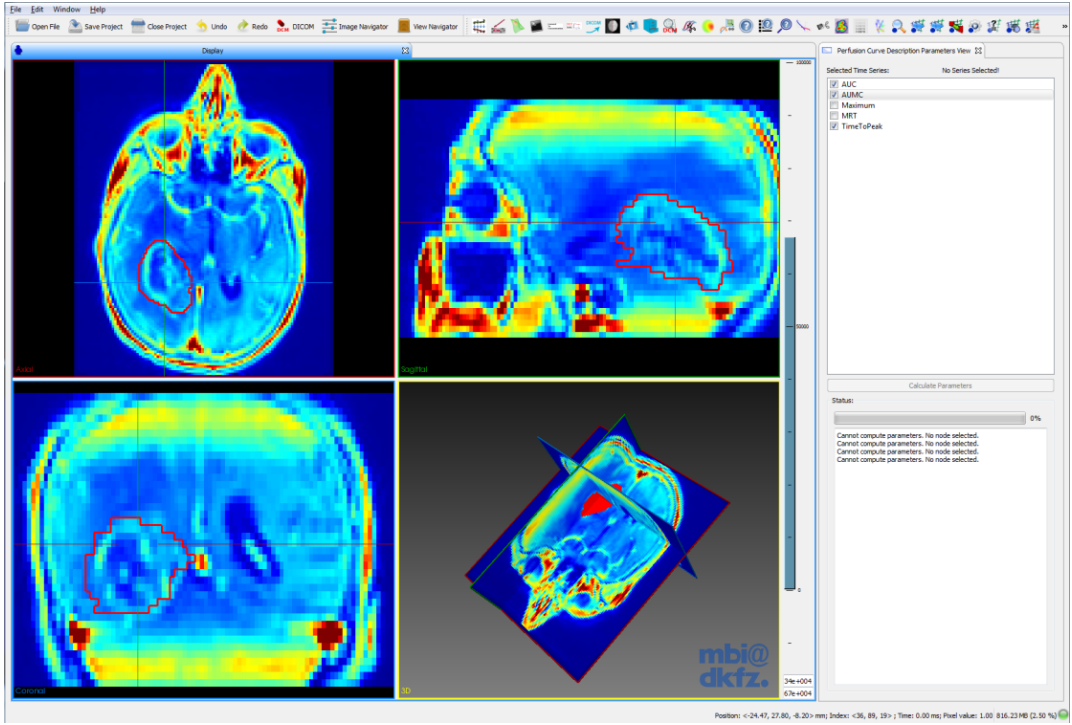

**Figure 6:** Screenshot of the MITK workbench and the curve description parameters plug-in that enables calculation of several semi-quantitative parameters, like the area-under-the-curve (AUC), time-to-peak and maximum signal. The images show the AUC calculated from the 4D DCE MRI data of a glioblastoma patient

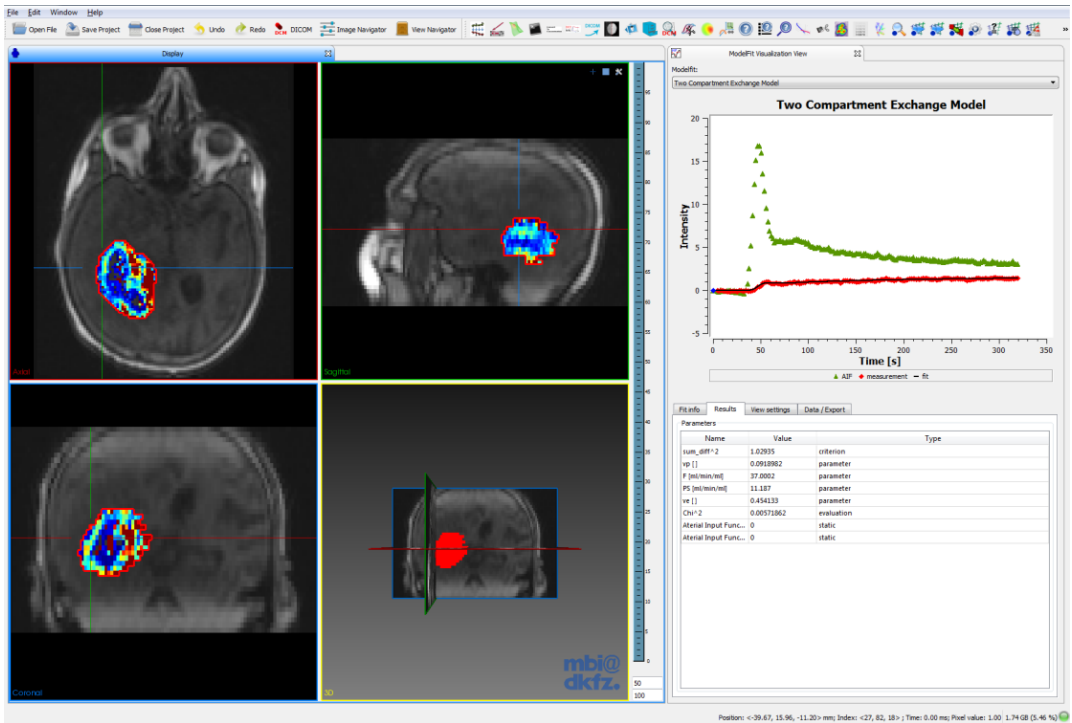

**Figure 5 :** Example of pharmacokinetic fitting analysis with the presented plug-in in a glioblastoma patient using the 2CXM within the tumor ROI (red). The 4-window view shows the first time frame of the dynamic MRI series, overlaid with the parameter map of plasma flow $F_p$. The MFI (right) shows the measured concentration-time curve in a single voxel (red dots), together with the estimated model fit (black line) and the used AIF (green dots). The respective model parameter estimates of the fit are listed in the table below

The plug-in comprises several options shown in Fig. 7. The AIF, required as input for these models, can be defined in different ways. Image-based AIFs can be defined through segmentation of a feeding artery and are then extracted from dynamic images, equivalent to the tissue concentration-time curves. The segmentation can be defined on the same dynamic image as the tissue of interest or on any other dynamic image. This feature is especially useful in preclinical studies, where usually a slice through the heart is acquired separately and used for derivation of the CA concentration in the blood pool. Another option to provide an AIF is via an external file in .csv format, which can be used e.g. for population-derived input curves [51]. Initial parameter values can be defined for each individual model parameter, either as a constant global value for all voxels or locally in form of a parameter image. Default values for the respective model are natively set. Constraints can be imposed on the model parameters, in order to exclude unrealistic values and to limit the search space. Upper and lower constraint values can be defined individually for each model parameter. Combinations, such as sums of parameters, are also possible. The tool allows limitation of the fitting region by definition of a segmentation for the region/volume of interest. Within this ROI, fitting can be performed in each individual voxel (voxel-wise) or on the averaged curve (ROI-based). Parameter estimates of the respective model, together with the used fitting criterion (e.g. the sum of squared residuals), are displayed in parameter images. Individual fit curves can be assessed using the MFI.

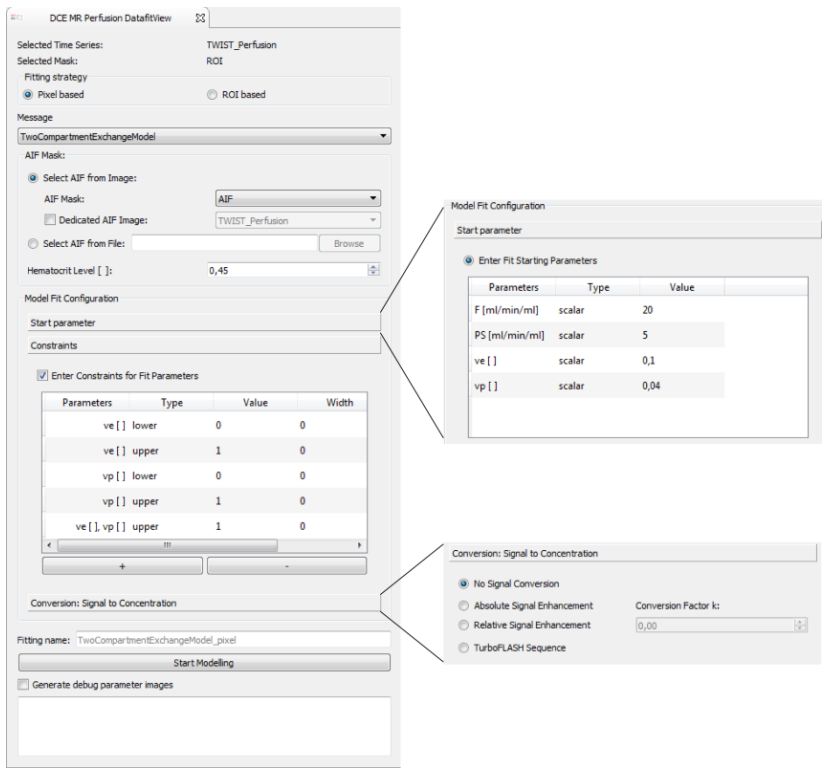

**Figure 7:** The DCE MRI analysis plug-in. Several models are currently implemented: a simple three-step linear function, a descriptive model, the standard and extended Tofts model and the 2CXM. Arterial input functions can be image-based (from the same image as the analyzed ROI or a different one) and file-based (.csv format). The fit configuration allows for definition of model parameter starting values, parameter constraints and the desired conversion of signal to concentration units

There is also an option to extract certain debug parameters describing technical statistics of the fitting process, such as the required optimization time, the number of fit iterations, or the convergence criterion. These are useful for evaluation of fit quality beyond standard criterion parameters and visual fit assessment, especially in cases of failed or non-terminated fits.

Before fitting can be performed, 4D DCE MRI image intensities usually need to be converted into the corresponding CA concentration. If pre-contrast $T_1$ maps are available (e.g. from multiple flip-angle measurements), analytic conversion of the signal to concentration units is provided in the DCE MRI fitting tool (as described in [27]). Otherwise, conversion by means of relative and absolute signal enhancement can be performed. The conversion can be performed in a dedicated plug-in or as convenient alternative directly in the fitting plug-in.

The versatility of our framework enabled also the implementation of a tool for simulating concentration-time curves by forward calculation of the signal from parameter images in combination with an AIF. With this simulation tool, curves can be generated according to the standard Tofts Model, the extended Tofts model and the 2CXM. Noise in form of Gaussian random numbers can be added with user-defined contrast-to-noise ratios (defined as ratio between the maximum of the AIF and the standard deviation of the noise). The generation of synthetic concentration-time curves allows for validation of models and benchmarking of different configurations of fitting algorithms [42].

For tracer-kinetic analysis of dynamic PET images, a dedicated tool was implemented in analogy to the DCE MRI tool. It includes the one tissue compartment model (1TCM) and the two tissue compartment model (2TCM). The first is provided in the general three-parameter form and a simplified 2 parameter version, while the latter is provided in a general form and in an adapted form for FDG PET, the most commonly used tracer [52, 53]. The general 2TCM function is provided in both the convolution and differential equation form. The fit options (e.g. types of AIF, initial parameters, constraints) are similar to the presented options for DCE MRI fitting (see Fig. 7). Conversion of signal intensities from raw data time-activity curves (TAC) to standard uptake value (SUV) curves can be performed using a separate plug-in. Figure 8 shows an example case for the $^{18}$F-labeled fluoroethyl-tyrosine (FET) tracer, which is commonly used for detection and staging of brain tumors [54, 55]. Time-activity curves were fitted with the standard 1TCM. Parameter maps of the exchange rates are shown together with fitted curves in a representative voxel.

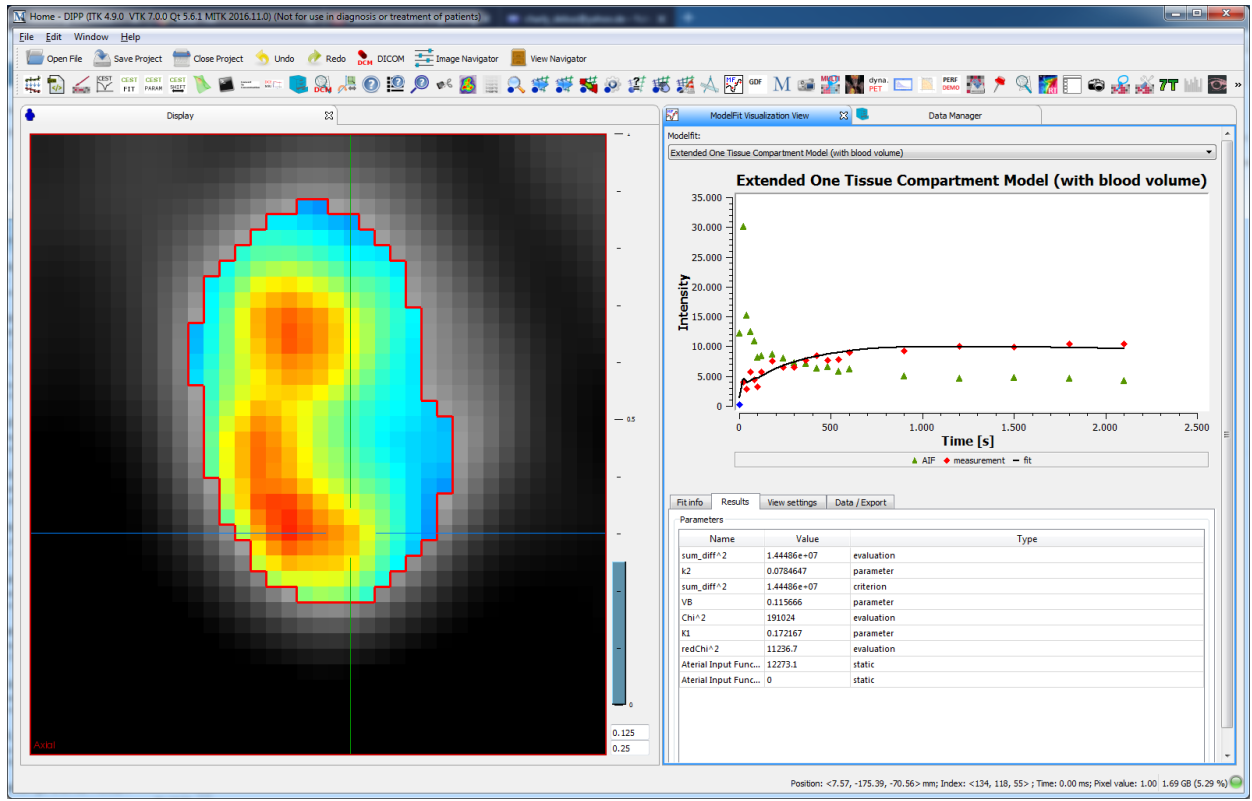

**Figure 8:** Example case of application of the fitting software for tracer-kinetic analysis in dynamic PET, using the tracer kinetic one tissue compartment model. The image shows the parameter map of the exchange rate $K_1$. In the MFI, the measured time-activity curve in tissue (red) with corresponding fit (black) and the utilized arterial time-activity curve (green) are displayed

A number of additional analysis tools for other workflows and fitting domains (e.g. for fitting of Z-spectra in chemical exchange saturation transfer (CEST) MRI, general $T_1$/$T_2$ mapping) were further derived from the fitting framework.

The above presented ready-to-use analysis tools are integrated as plug-ins into the MITK workbench and can be run directly via the user interface. In addition, they can be used as command-line tools (CLI tool) for (semi-) automatic analysis, which makes efficient evaluation of large data cohorts feasible. The CLI tools are implemented using the CommonTK/Slicer execution model [56] and therefore offer an easy integration path into other applications without any compile or linkage dependencies.

The integration into MITK allows establishing a complete analysis pipeline. Pre- and post-processing of raw data and resulting parameter maps can be done using all available MITK functionalities. Dynamic images can be co-registered with other, static images of higher signal-to-noise ratio (SNR) or spatial resolution, which enables more precise lesion detection and subsequent segmentation. Manual and semi-automated image segmentation techniques facilitate definition of fitting ROIs and other inputs (e.g. AIF). Segmentations can then be directly used for voxel-based as well as ROI-based fitting without further data conversion. Generated model parameter maps from fitting or non-compartmental analysis can be handled independently as MITK images, and thus analyzed. Besides visual inspection of the individual fit values and curves using the MFI plug-in, statistics and histogram evaluations can be performed. Further segmentations of sub-regions can be derived for in-depth analysis. Parameter maps can be saved in various image formats or exported to .csv files for further analysis with other programs.

## Validation

Validation was performed for the standard and the extended Tofts model as well as the 2CXM. The quantitative imaging biomarker alliance (QIBA) offers virtual phantom data, so-called digital reference objects (DRO), for pharmacokinetic analysis in DCE MRI. For validation of the standard 2-parameter Tofts model, the QIBA_v6_Tofts DRO was used[†]. It contains 30 blocks of each 10 × 10 squares of combinations of $K_{trans} \in \{0, 1, 2, 5, 10, 20, 35\}$ ml/min/100 ml and $v_e \in \{0.01, 0.05, 0.1, 0.2, 0.5\}$. Concentration curves are sampled at a temporal resolution of $0.5$ s over a total of 1361 time points. For validation of the extended 3-parameter Tofts model, the noise free 4D DRO[‡] was chosen. It includes DICOM images of a 2D+t DCE MRI series, with each 10 x 10 voxel blocks of 108 different, spatially encoded concentration-time curves, using all combinations of $K_{trans} \in \{0, 1, 2, 5, 10, 20\}$ ml/min/100 ml, $v_p \in \{0.001, 0.005, 0.01, 0.02, 0.05, 0.1\}$ and $v_e \in \{0.1, 0.2, 0.5\}$. Concentration curves are sampled at a temporal resolution of $0.5$ s over a total of 661 time points. The dataset was fitted with the DCE MRI tool and resulting parameter estimates were compared to the ground truth. Unfortunately, no DRO is available for the 2CXM to our knowledge. Thus, we created a third DRO for the 2CXM, similar to those for the standard and extended Tofts model. Concentration-time curves for different combinations of $F_p$, $PS$, $v_p$ and $v_e$ were generated from the 2CompFlowExch model in JSim [57]. Values used for generation of concentration-time curves were $F_p \in \{5, 10, 25, 40\}$ ml/min/100 ml, $PS \in \{0.0, 5, 15\}$ ml/min/100 ml , $v_p \in \{0.02, 0.05, 0.1, 0.2\}$ and $v_e \in \{0.1, 0.2, 0.5\}$. Curves were simulated on $0.5$ s temporal sampling using the arterial input curve extracted from the extended Tofts DRO data. Signal curves were arranged in a 3D+t image of spatial dimensions $< 10 \cdot F_p, 10 \cdot v_e \cdot v_p, PS > = < 40, 120, 3 >$, where each $< 10 \times 10 \times 1 >$ voxels contained one curve type. The AIF was added as $< 40, 20, 3 >$ block on the bottom of the image, leading to a final DCE MRI data set of dimension $< 40, 140, 3, 661 >$. The data is available in DICOM format under: http://mitk.org/wiki/MITK-ModelFit. All three DROs of synthetic DCE MRI data were fitted with our model implementation, using the AIF within the images, and resulting parameter estimates were compared to the true values. Mean relative errors on parameter estimates are listed in Table 2. For the 2CXM, errors are subdivided into different cases of $PS$.

---

[†] https://sites.duke.edu/dblab/qibacontent/
[‡] https://sites.duke.edu/dblab/dce-mri-test-images/

The standard Tofts model yielded mean errors of 3.24 % $\pm$ 0.97 % for $K_{trans}$ and 0.38 % $\pm$ 1.99 % for $v_e$, ranging between $-7.44$ % and $+2.22$ % (error maps not shown). Figure 9 shows relative errors on parameter estimates $K_{trans}$, $v_p$ and $v_e$ for the extended Tofts model. Largest errors on $K_{trans}$ and $v_e$ were observed for lowest $K_{trans}$ values of 1 ml/min/100 ml. $v_p$ exhibited largest errors for $v_p = 0.001$. These findings are reasonable, since perfusion is difficult to measure in cases with low overall perfusion (low $K_{trans}$) and low vascularization (low $v_p$). Apart from these cases, errors on parameter estimates were low, between 0% and 10% in most cases.

Figure 10 shows relative errors on $F_p$ and $v_p$, for each of the three different original values of $PS = 0, 5, 15$ ml/min/100 ml. Original values of $v_p$, $v_e$ and $F_p$ are indicated on the axes. These 2D error maps were generated by averaging each 10 slices with respective $PS$ values. Estimates on $F_p$ presented with very low errors of approximately 2% on average. Largest errors of about 4% are found for $PS = 0$ ml/min/100 ml, $v_p = 0.1$ and $F_p = 40$ ml/min/100 ml. For estimates on $v_p$, largest errors were found for $PS = 0$ ml/min/100 ml, at $v_p = 0.1$ and $F_p = 25$ ml/min/100 ml.

|  |  |  | Mean Error [%] |
|---|---|---|---|
| Standard Tofts | $K_{trans}$ |  | 3.24 |
|  | $v_e$ |  | 0.38 |
| Extended Tofts | $K_{trans}$ |  | 7.01 |
|  | $v_p$ |  | 22.06 |
|  | $v_e$ |  | 5.95 |
| 2CXM | $F_p$ | $PS = 0\ ml/min/100\ ml$ | 2.54 |
|  |  | $PS = 5\ ml/min/100\ ml$ | 1.99 |
|  |  | $PS = 15\ ml/min/100\ ml$ | 1.86 |
|  | $v_p$ | $PS = 0\ ml/min/100\ ml$ | 10.55 |
|  |  | $PS = 5\ ml/min/100\ ml$ | 1.83 |
|  |  | $PS = 15\ ml/min/100\ ml$ | 2.87 |
|  | $PS$ | $PS = 5\ ml/min/100\ ml$ | 3.05 |
|  |  | $PS = 15\ ml/min/100\ ml$ | 2.69 |
|  | $v_e$ | $PS = 5\ ml/min/100\ ml$ | 3.22 |
|  |  | $PS = 15\ ml/min/100\ ml$ | 2.48 |

**Table 2:** Relative Errors on parameter estimates from fitting the validation datasets.

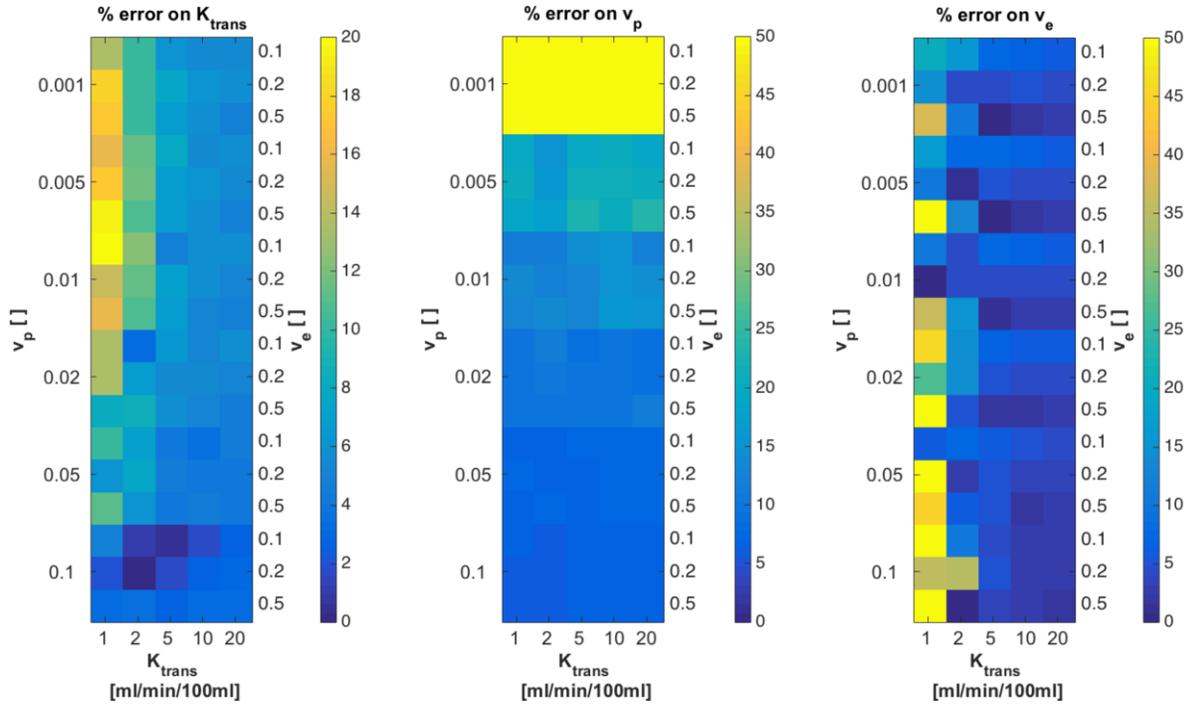

**Figure 9:** Relative errors on $K_{trans}$, $v_p$ and $v_e$ from fits with our implementation of the extended Tofts model to the noise free 4D QIBA digital reference object. True parameter values, used to create the DRO, are indicated on the left ($v_p$), right ($v_e$) and bottom ($K_{trans}$) scale, in order to see patterns of errors for certain tissue types

Overall errors on $v_p$ for the other two cases of $PS = 5, 15 \text{ ml/min}/100 \text{ ml}$ were around $2\% - 3\%$. For analysis of estimates on $PS$ and $v_e$, cases with $PS = 0 \text{ ml/min}/100 \text{ ml}$ were excluded, as large errors are to be expected for these two parameters in cases with vanishing vascular permeability. For $PS = 5 \text{ ml/min}/100 \text{ ml}$ and $PS = 15 \text{ ml/min}/100 \text{ ml}$, resulting errors on parameter estimates for $PS$ and $v_e$ are presented in Fig. 11. Errors on both $PS$ and $v_e$ were about 3% on average, except for $PS = 5 \text{ ml/min}/100 \text{ ml}$, $F_p = 5 \text{ ml/min}/100 \text{ ml}$ with $v_p = 0.2$ and 0.02 and $v_e = 0.1$ and 0.5, respectively. All these findings are reasonable, since limit tissues (low $F_p$ in combination with high $v_p$, low $PS$ in combination with high $v_e$) are expected to yield larger errors, as correct determination of concentration-time curves is difficult and assumptions of the 2CXM are not entirely valid in these cases.

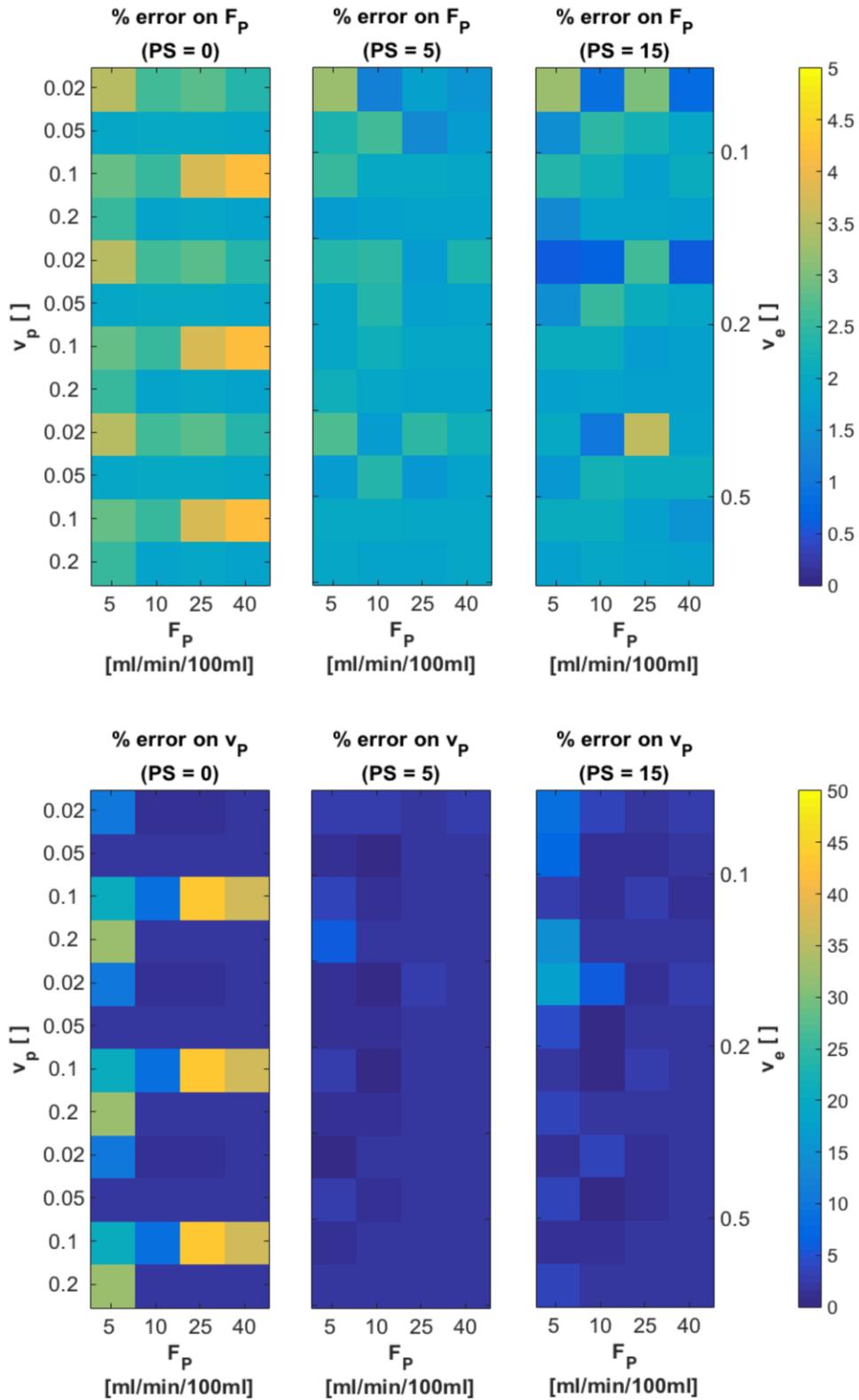

**Figure 10:** Relative errors on parameter estimates of $F_p$ and $v_p$ for three cases of $PS = 0$ *ml/min/100 ml*, $PS = 5$ *ml/min/100 ml* and $PS = 15$ *ml/min/100 ml* from fits with our implementation of the 2CXM to the digital reference object created using JSim. True parameter values, used to create the DRO, are indicated on the left ($v_p$), right ($v_e$) and bottom ($F_p$) scale, in order to see patterns of errors for certain tissue types

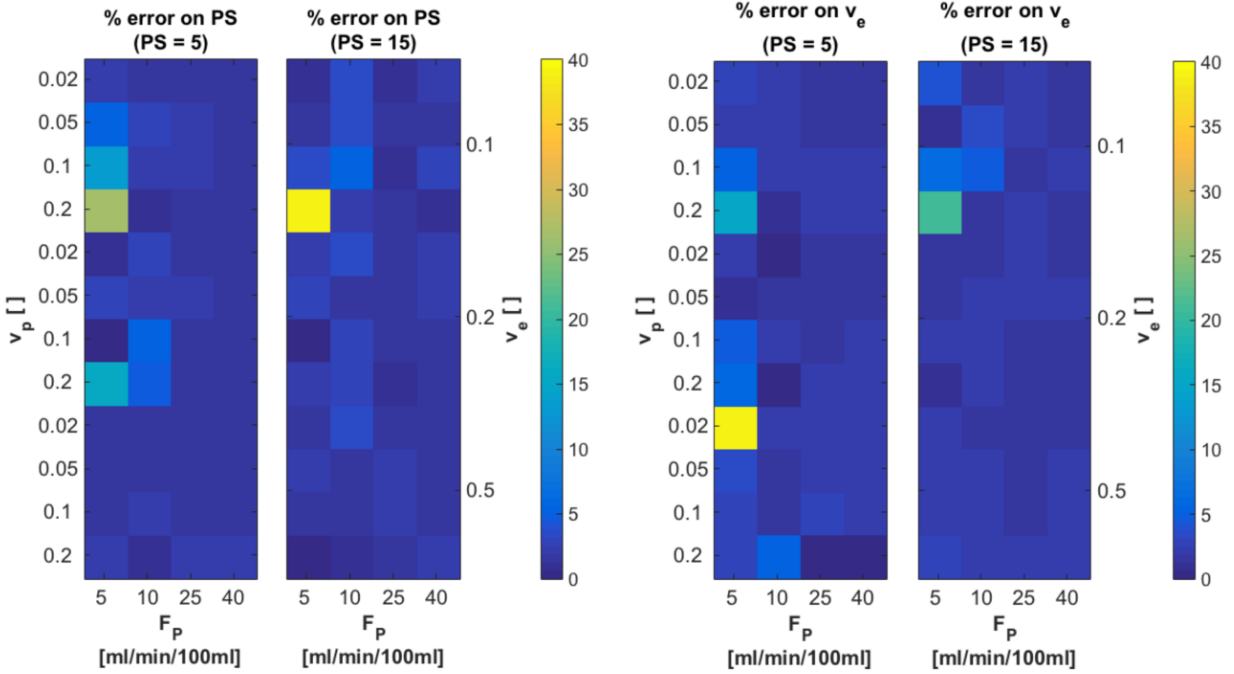

**Figure 11:** Relative errors on parameter estimates of *PS* and $v_e$ for two cases of *PS* = 5 *ml/ min* /100 *ml* and *PS* = 15 *ml/ min* /100 *ml* from fits with our implementation of the 2CXM to the digital reference object created using JSim. True parameter values, used to create the DRO, are indicated on the left ($v_p$), right ($v_e$) and bottom ($F_p$) scale, in order to see patterns of errors for certain tissue types

# Discussion

We presented an open-source software framework for fitting of medical images within the Medical Imaging Interaction Toolkit MITK. Its implementation offers high flexibility and reusability, making it easily adaptable and extendable for own developments. This versatility enables development of analysis tools for various fitting tasks and image modalities, as for example fitting of Z-spectra in CEST MRI or tracer-kinetic analysis in dynamic PET using compartment models. Furthermore, means of qualitative and quantitative fit quality evaluation and result visualization are provided.

Ready-to-use applications in form of plug-ins for the MITK workbench were implemented for several dedicated use-cases, which allows for direct GUI-based analysis. An extensive toolbox for pharmacokinetic analysis of DCE MRI data was designed with several commonly used pharmacokinetic models. It offers a wide range of configuration options, such as definition of parameter starting values, constraints for model parameters and different methods to convert the acquired MR signal to contrast agent concentration. Several strategies for AIF definition are supported, in order to enable most common approaches, e.g. image-based or population-based AIFs. Fitting can be performed either on an individual voxel basis or in a ROI-based average approach. The most commonly used models "standard Tofts", "extended Tofts" and "2CXM" were validated on digital reference data sets. Results of estimated parameters for the standard and extended Tofts model were comparable to other published data [16, 27]. For the 2CXM we created a synthetic data set using JSim [57]. In order to spread validation methods and thus standardize pharmacokinetic modeling in DCE MRI, we provide free access to our validation DRO for the 2CXM.

Many approaches in medical image analysis utilize fitting, from pre-processing over extraction of parameters as in pharmacokinetic analysis to simple modeling of time-dependent treatment effects. Commonly, image fitting is performed using in-house developed code in large data analysis platforms such as MATLAB, IDL/MPFIT or R [13, 14, 15]. For analysis of DCE MRI data using pharmacokinetic modeling, several groups have presented open-source software solutions. Even though these tools are made publicly available as open-source code, several (e.g. PMI based on IDL or ROCKETSHIP based on MATLAB) depend on commercial software. Smith et al. presented DCEMRI.jl [27], a toolkit for NLLS fitting analysis of DCE MRI written purely in the programming language *julia*, hence it does not depend on any commercial licenses. It is open-source and compatible with MacOS, Linux and Windows systems. However, no graphical user interface or fit visualization is provided. Furthermore, it is not incorporated into any image processing platform, and input and output of data is possible only in MAT-v5 files. Hence a clinical oriented workflow, depending on the support of DICOM formats, is very difficult to realize. Published packages for R or Python, for example DATforDCEMRI [25], depend on no commercial licenses, as these frameworks are freely available and applicable to all operating systems. However, they are not dedicated to medical image analysis, thus data import and integration into a DICOM workstation are not provided. Additionally, other image processing tasks, like segmentation or registration, have to be performed externally. The OsiriX [33] open-source DICOM workstation can be used for an entire radiological workflow. Thus, OsiriX plug-ins for DCE MRI analysis, e.g. UMMPerfusion [31], are popular tools in the context of clinical research. However, OsiriX is only available on Mac OS, which presents another drawback. Furthermore, these plug-ins are specifically designed and implemented for OsiriX, hence they cannot be used as stand-alone tools, or with in automated, batch-processing analysis pipelines. 3DSlicer offers a rich, open-source and OS-independent platform for medical image analysis and visualization. The PKModeling module [35] can be used in automated workflows or via the 3DSlicer GUI. The major drawbacks compared to other options are the limited number of available models.

Our framework overcomes these limitations, by offering a standardized software concept for data handling, fitting algorithmic and analysis pipelines that can be applied modularly and extended easily. The level of abstraction, compared to other solutions, does not stop with the introduction of normal MVC patterns to separate GUI and algorithms. The fitting infrastructure itself is carefully abstracted and standardized. This ensures a large degree of freedom with respect to both the use-case in question (image modality, fitting domain) as well as the specific algorithm configuration in a use-case (optimizer and cost function, model). Due to the toolkit nature of MITK and the framework, the implemented tools for specific use-cases (DCE MRI, dynamic PET, general purpose fitting) can be used for automated batch processing, for integration into other applications (e.g. as CLI tool), as well as for direct user-interaction with GUI-based applications (plug-ins for the MITK workbench) for end-users without need for advanced software development. Furthermore, the embedding in MITK allows for fitting to be performed within an eco-system of medical image processing combining all other relevant processing steps and without the burden of data conversion or inter-application transfer. This embedding, together with the capability of MITK to handle all different kinds of imaging modalities, is especially useful in the light of current efforts regarding modern hybrid PET-MRI scanners or other kinds of multi-parametric imaging data. Different dynamic data types can be evaluated side-by-side in the same software framework and thus allow direct comparison of the derived multi-parametric maps. MITK also enables the analysis of longitudinal data generated by treatment response studies like in neoadjuvant chemotherapy, radiation therapy or anti-angiogenic treatment. The data analysis tools of ModelFit can be utilized to evaluate the physiologic, metabolic and vascular status of the tumor tissue and thus assess treatment efficacy, as changes in parametric maps reflect responses to therapy. Additionally, the presented software framework can be considered as truly free and open-source as it requires neither further proprietary licenses nor is it limited to specific operating systems. These features also facilitate the wide spread use of the implemented tools and thus can aid in standardization and multi-center analyses.

The quantitative imaging biomarker alliance (QIBA) [58] aims to reduce variability of quantitative imaging biomarkers across devices, sites, patients and time, and thus improve their value and practicality. In recent years, substantial efforts have been made to include, amongst others, pharmacokinetic approaches in MRI and PET into the alliance through standardized validation datasets, software approaches and acquisition protocols. Within this context, our framework for fitting of medical image data could provide a huge step forward in standardizing software not only for DCE MRI, as it can provide a common basis for all application of fitting approaches, whilst being freely available, maintained and transparent (i.e. source code can be directly accessed).

To our knowledge there is no software package for DCE MRI pharmacokinetic analysis that (1) has more functionalities regarding pharmacokinetics (2) can be considered truly free as in open-source and operating system independent, (3) with both GUI and batch processing applicability, (4) integration into a global image processing platform for pre- and post-processing and (5) no dependencies on commercial software packages or licenses.

# Conclusion

In summary we have designed and implement a highly flexible and easily extendable software environment for fitting analysis in medical imaging, that allows for the analysis to be performed directly integrated into a larger image pre- and post-processing workflow. This includes, amongst others, highly automated evaluation workflows regarding longitudinal data, where modeling of responses (e.g. follow-up under therapy) comes into play and which has become increasingly important in the era of big data. The framework can be used by developers for custom developments, but also offers ready-to-use GUI based applications for end-users. It is open-source and OS-independent, which together with its high modularity, versatility and rich feature set makes it superior to other existing solutions, especially in the context of pharmacokinetic analysis of dynamic imaging data.

# Declarations

### Funding
This work was supported by the National Center for Tumor diseases (NCT 3.0-2015.22 BioDose, to AA) the German Research Foundation (DFG-KFO-214, to AA; and SFB/TRR 125 "Cognition-Guided Surgery", to KMH and MN), the Federal Ministry of Education and Research Germany (BMBF 01IB13001B, to RF) and the Deutsche Krebshilfe (Max-Eder 108876, to AA). The funders had no role in study design, data collection and analysis, decision to publish or preparation of the manuscript.

### Authors' contributions
CD, RF, MN, KMH and AA participated in design and supervision of the project. CD and RF implemented the software. CD analyzed the data. MI and IK provided background knowledge on DCE MRI. CD and MI collected patient data. MN and KMH supplied the necessary software support for MITK. CD and RF wrote the manuscript. All authors read and approved the final version of the manuscript.

### Acknowledgements
The authors would like to thank Alina Bendinger, Christin Glowa, Maria Saager, Christian Karger, Patrick Schünke, Jennifer Mosebach, David Bonekamp, Patrick Badura and Dorde Komljenovic, for fruitful discussions on perfusion imaging and pharmacokinetic analysis, and testing of developed plug-ins. Furthermore we acknowledge Caspar Goch, Stefan Dinkelacker and the division of medical image computing for continuous software support and Ali Afshar and Uwe Haberkorn for insight knowledge on dynamic PET data evaluation.

# Availability and requirements

**Project name:** MITK ModelFit
**Project home page:** http://mitk.org/wiki/MITK-ModelFit
**Operating system(s):** Platform independent
**Programming language:** C++14
**Other requirements:** Qt 5.9 or higher, Cmake 3.10 or higher; Git from http://git-scm.com
**License:** BSD-like
**Any restrictions to use by non-academics:** none

# List of abbreviations

| | |
|---|---|
| AIF | Arterial input function |
| AUC | Area-under-the-curve |
| CEST | Chemical exchange saturation transfer |
| CLI | Command line interface |
| CNR | Contrast-to-noise ratio |
| DRO | Digital reference object |
| GUI | Graphical user-interface |
| MITK | Medical Imaging Interaction ToolKit |
| MP | Model parameters |
| MFI | Model-fit inspector |
| MVC | Model-View-Controller |
| QIBA | Quantitative Imaging Biomarker Alliance |
| ROI | Region of interest |
| SMP | Static model parameters |
| SNR | Signal-to-noise ratio |
| SUV | Standard uptake value |
| TAC | Time-activity curve |
| TTP | Time-to-peak |
| 1TCM | One-tissue compartment model |
| 2CXM | Two compartment exchange model |
| 2TCM | Two-tissue compartment model |